\documentclass[fleqn,12pt,twoside]{article}
\usepackage{espcrc1}
\usepackage{epsf,graphics,rotate}

\newcommand{\insertfig}[2]{\mbox{\epsfxsize=#1cm \epsfbox{#2.eps}}}
\newcommand{\Bx}{x_{\rm B}}

\newcommand{\cQ}{{\cal Q}}
\newcommand{\GeV}{\mbox{GeV}}

\newcommand{\AmS}{{\protect\the\textfont2
  A\kern-.1667em\lower.5ex\hbox{M}\kern-.125emS}}

\title{Generalized Parton Distributions -- theoretical review --}

\author{D. M\"uller
\address{Institut f\"ur Theoretische Physik, Universit\"at  Regensburg,
D-93040 Regensburg, Germany} }

\begin{document}

\thispagestyle{empty}
\vspace{5mm}

\centerline{\large \bf Generalized Parton Distributions}
\centerline{\large \bf -- theoretical review --}

\vspace{10mm}
\centerline{\bf D. M\"uller}

\vspace{5mm} \centerline{\it Institut f\"ur Theoretische Physik}

\centerline{\it Universit\"at Regensburg}

\centerline{\it D-93040Regensburg, Germany}

\vspace{15mm}

 \centerline{\bf Abstract}

 \vspace{5mm}
\noindent
In this talk an introduction to generalized parton
distributions is given. Recent developments are shortly reviewed,
including non-perturbative calculations, phenomenological aspects
and evaluation of higher order perturbative and power corrections.

\vspace{120mm}

\centerline{\it Invited talk given at the international
conference} \centerline{\it Baryons 2004 in Palaiseau, France,
25.-29.\ October  2004.} \setcounter{page}{0}

\maketitle

\begin{abstract}
In this talk an introduction to generalized parton distributions
is given. Recent developments are shortly reviewed, including
non-perturbative calculations, phenomenological aspects and
evaluation of higher order perturbative and power corrections.
\end{abstract}

\section{Introduction}

Generalized parton distributions (GPDs)
\cite{GeyDitHorMueRob88MueRobGeyDitHor94,Ji96,Rad96,ColFraStr96}
and their crossed version, i.e., the generalized distribution
amplitudes (GDAs)
\cite{GeyDitHorMueRob88MueRobGeyDitHor94,DieGouPirTer98} appear in
the perturbative description of certain hard exclusive processes,
e.g., in the leptoproduction of photon and mesons or in the
production of hadron pairs in two photon annihilation. Although
the first systematic study of GPDs and GDAs was done for more than
one decade \cite{GeyDitHorMueRob88MueRobGeyDitHor94}, their
physical significance has been widely realized in connection with
the proton spin puzzle. Namely, these distributions incorporate,
besides other non-perturbative information that is not encoded in
forward parton densities or form factors, also gravitational form
factors from which the quark orbital angular momentum fraction
contributing to the nucleon spin can be read off \cite{Ji96}. This
perception induced then intensive theoretical and experimental
studies, in which it has been fully realized that GPDs and GDAs
are a new concept to study the structure of hadrons and nuclei,
contain a link between exclusive and inclusive processes and open
a new window for the exploration of non-perturbative QCD. In the
following a mini review about these developments is given, for
comprehensive ones see Refs.\
\cite{Ji98,Rad00,GoePolVan01,Die03a}.

In Sect.~\ref{Sec-FeaGenParDis} the basic properties of GPDs and
their partonic interpretation are presented. In
Sect.~\ref{Sec-ModLatCal} GPD ans\"atze are shortly discussed  and
is then devoted to the non-perturbative evaluation. The exclusive
processes, which allow to gain access to these distributions, are
listed in Sect.~\ref{Sec-ExcReaLO}. In Sect.~\ref{Sec-BeyLO}
results from the evaluation of radiative and power suppressed
corrections beyond leading order (LO) are reported and finally,
the conclusions are given.

\section{Features of Generalized parton distributions}
\label{Sec-FeaGenParDis}

GPDs are defined as Fourier transform of light-ray operators,
sandwiched between the initial and final hadronic states. There is
a whole compendium of GPDs: corresponding to the species of
hadrons, the initial and final states can have different quantum
numbers (transition GPDs), and even we might replace the hadrons
by nuclei (nucleus GPDs). Specified the initial and final states,
GPDs are classified with respect to the twist of the operators and
the spin content of fields. At leading twist-two level three
different types of quark and gluon GPDs can be defined (here the
gauge link is omitted):
\begin{eqnarray}
\label{Def-GPD-q}
\left\{\!\!\!
\begin{array}{c}
 {^q\!q}^V \\ {{^q\!q}^A }  \\ {{^q\!q}^T }
\end{array}
\!\!\!\right\}(x,\xi,\Delta^2,\mu^2) \!\!\! &=&\!\!\! \int\!
\frac{d\kappa}{2\pi}\; e^{i \kappa { x} P_+}
     \langle P_2, S_2 \big|\bar{\psi}_q^r(-\kappa n)
\left\{\!\!\!
\begin{array}{c}
{ \gamma_+}\\ {\gamma_+\gamma_5} \\ {i\sigma_{+\perp}}
\end{array}
\!\!\!\right\}\psi^r_q(\kappa n) \big|P_1, S_1 \rangle ,
\\
\label{Def-GPD-g}
\left\{\!\!\!
\begin{array}{c}
{ {^G\!q}^V} \\ {{^G\!q}^A} \\   { {^G\!q}^T}
\end{array}
\!\!\!\right\}(x,\xi,\Delta^2,\mu^2) \!\!\! &=&\!\!\!  2
\int\!\frac{d\kappa}{\pi P_+}\;  e^{i \kappa {x} P_+}
         \langle P_2,S_2 \big| G^a_{+\mu}(\!-\kappa n)\!\!
\left\{\!\!\!
\begin{array}{c}
{ g_{\mu\nu}}\\ { i\epsilon_{\mu\nu-+}} \\ {
\tau_{\mu\nu;\rho\sigma}}
\end{array}
 \!\!\!\right\}\!\! G^a_{\nu+}(\kappa n)\big| P_1,S_1 \rangle ,
\end{eqnarray}
with $P_+ = n\cdot (P_1+P_2), V_- = n^\ast \cdot V,\ n^2=
(n^\ast)^2=0, n\cdot n^\ast=1$. In the first (vector) and second
(axial-vector) entry the in- and outgoing partons have the same
helicities, where its sum or difference of left- and right-handed
partons is taken, respectively. For the third entry, called
transversity, a helicity flip appears. GPDs depend on the momentum
fraction $x$, conjugated to the light-cone distance $2\kappa$, the
longitudinal momentum fraction $\xi = (P_1 -P_2)^+/(P_1 +P_2)^+$
in the $t$-channel, the momentum transfer $\Delta^2\equiv t=(P_2
-P_1)^2$, and the renormalization scale $\mu^2$. The latter is
induced by the renormalization prescription of the operators,
which is  part of the GPD definition. To deal with the
polarization of the hadronic states, one might introduces a form
factor decomposition \cite{Ji98,BerCanDiePir01}. For instance, for
the nucleon GPD Dirac and Pauli-like form factors appear in the
vector case \cite{Ji98}:
\begin{eqnarray}
\label{Def-ForFacDec}
{^i\!q}^V= \overline{U}(P_2,S_2)  \gamma_+
U(P_1,S_1) {H_i}(x,\xi,\Delta^2) + \overline{U}(P_2,S_2)
\frac{i\sigma_{+\nu} \Delta^\nu}{2 M} U(P_1,S_1)
{E_i}(x,\xi,\Delta^2)\,,
\end{eqnarray}
where $i=u,d,s,\cdots, G$. Definitions
(\ref{Def-GPD-q})--(\ref{Def-ForFacDec}) imply the basic
properties:
\begin{itemize}
\item
GPDs reduce  in the forward limit $\Delta\to 0$ to parton
densities
\cite{GeyDitHorMueRob88MueRobGeyDitHor94,Rad96,Ji96,ColFraStr96},
e.g.,
\begin{eqnarray}
q_i(x,\mu^2) = \lim_{\Delta\to 0} {H_i}(x,\xi,\Delta^2, \mu^2)\,,
\quad\mbox{and} \quad \lim_{\Delta\to 0} {E_i}(x,\xi,\Delta^2,
\mu^2) \not= 0,
\end{eqnarray}
while the helicity flip contribution ${E_i}$ decouples.
\item
 The $\mu^2$-dependence is governed by linear evolution equations
\cite{GeyDitHorMueRob88MueRobGeyDitHor94},
 which can be derived from the renormalization group equation of
the light-ray operators \cite{BorRob80,Bal83}. \item Hermiticity
together with time reversal invariance leads to a definite
symmetry with respect to the skewness parameter $\xi$, e.g.,
${H_i}(x,\xi)={H_i}(x,-\xi).$ \item The Mellin moments of GPDs are
expectation values of local twist-two operators:
\begin{eqnarray}
\label{Def-MellMom} \int\! dx\; x^n\; {^q\!q}^V(x,\xi,\Delta^2,
Q^2) = n^{\mu_0}\cdots n^{\mu_n} \langle P_2, S_2 \big| {\mbox{\bf
S}}\, \bar{\psi}_q^r { \gamma_{\mu_0}}\, i\!
\stackrel{\leftrightarrow}{D}_{\mu_{1}}\cdots  i\!
\stackrel{\leftrightarrow}{D}_{\mu_{n}} \psi^r_q  \big|P_1, S_1
\rangle\,.
\end{eqnarray}
Lorentz covariance induces  that they are polynomials in $\xi$,
which are even in (\ref{Def-MellMom}).
\end{itemize}
Furthermore,  GPDs are constrained in the region $x\ge |\xi|$ by
the positivity of the norm in the Hilbert space of states. The
most general form of such positivity bounds
\cite{MarRys97PirSofTer98,Rad98a}, known so far, are given as an
infinite set of constraints \cite{Pob02Pob03}. Since GPDs are
implicitly scheme dependent, positivity bounds are a request on
both the GPD model and scheme.  Note also that GPDs can be
represented as overlap of light-cone wave functions
\cite{DieFelJakKro00}.

\begin{figure}[t]
\begin{center}
\mbox{
\begin{picture}(600,90)(0,0)
\put(-10,0){\insertfig{5}{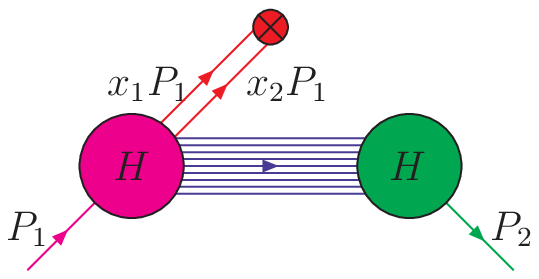}}
\put(140,0){\insertfig{5}{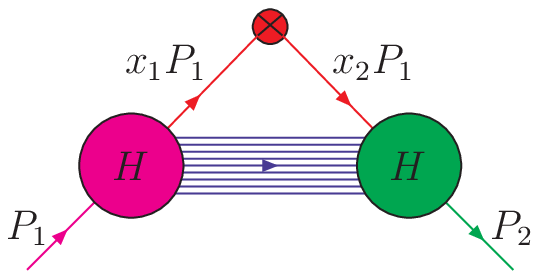}}
\put(310,-35){\insertfig{5}{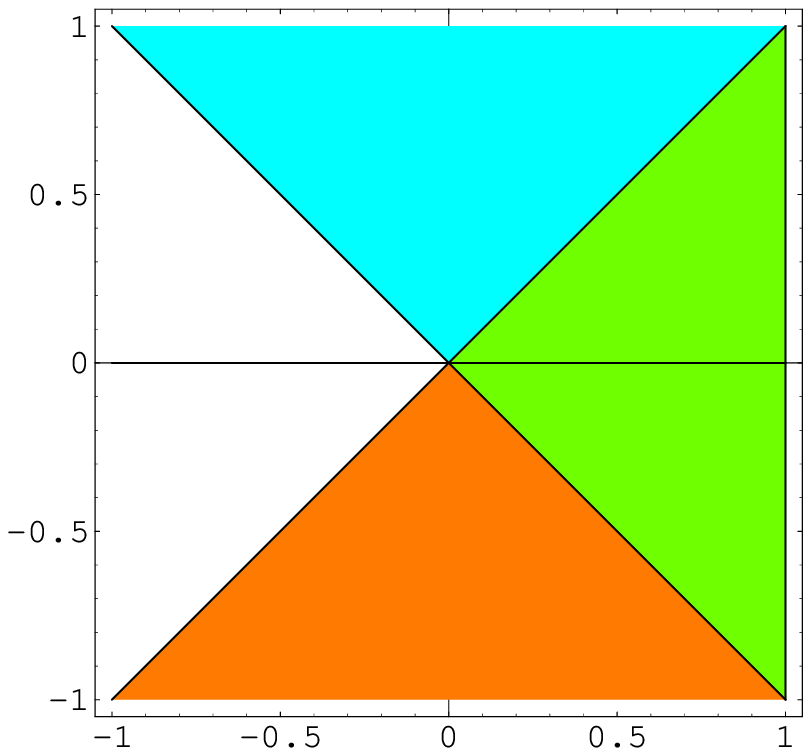}}
\put(375,70){$\scriptscriptstyle
\varpi\left(\frac{x}{\xi},\frac{1}{\xi}\right)$}
\put(418,65){$\scriptscriptstyle
\varpi\left(\frac{x}{\xi},\frac{1}{\xi}\right)$}
\put(400,48){$\scriptscriptstyle
-\varpi\left(\frac{-x}{\xi},\frac{-1}{\xi}\right)$}
\put(367,5){$\scriptscriptstyle
\varpi\left(\frac{-x}{\xi},\frac{-1}{\xi}\right)$}
\put(405,30){$\scriptscriptstyle
\varpi\left(\frac{-x}{\xi},\frac{-1}{\xi}\right)$}
\put(413,15){$\scriptscriptstyle
-\varpi\left(\frac{x}{\xi},\frac{1}{\xi}\right)$}
\put(450,-35){$\scriptstyle  x$} \put(310,95){$\scriptstyle \xi$}
\end{picture}
}
\end{center}
\caption{\label{FigParInt} Partonic interpretation  and support
property (right) of GPDs.}
\end{figure}
Let us give a partonic interpretation of GPDs. A generic quark
GPD, e.g., in the vector case, $q(x,\xi,\Delta^2)=
Q(x,\xi,\Delta^2) - \overline{Q}(-x,\xi,\Delta^2)$ is decomposed
in its quark $Q$ and anti-quark $\overline{Q}$ part. In the
central (exclusive or ER-BL) region $-\xi \le x \le \xi$,
$Q(x,\xi,\Delta^2)$ might be interpreted as probability amplitude
to have a meson like configuration inside the hadron, while the
outer (inclusive or DGLAP) region $\xi \le x \le 1$ can be viewed
as probability amplitude for emission  and absorbing a quark with
momentum fraction $x_1 P_1 = \frac{x+\xi}{1+\xi} P_1$ and $x_2 P_1
= \frac{x-\xi}{1+\xi} P_1$, respectively. Lorentz invariance ties
this both regions, which can be read off from the
representation\footnote{Eq.\ (\ref{Dec-GPD-1}), where
$\varpi(x,y,\Delta^2) = \int_0^{\frac{1+x}{1+y}}\!dw\, d(x,x-w
y,\Delta^2)$ might be derived by means of a partonic Fock state
decomposition and the so-called $\alpha$-representation for
Feynman diagrams. } that ensures polynomiality (valid for $\xi \ge
0$):
\begin{eqnarray}
\label{Dec-GPD-1} Q(x,\xi,\Delta^2)= \theta\left(-\xi \le x \le
1\right)
\frac{1}{\xi}\varpi\left(\frac{x}{\xi},\frac{1}{\xi},\Delta^2\right)
- \theta\left(\xi \le x \le 1\right) \frac{1}{\xi}
\varpi\left(-\frac{x}{\xi},-\frac{1}{\xi},\Delta^2\right)\,.
\end{eqnarray}
Obviously,  in the central region the GPD is given by
$\varpi\left(\frac{x}{\xi},\frac{1}{\xi}\right)$ from which the
outer region, determined by the antisymmetric function
$\varpi\left(\frac{x}{\xi},\frac{1}{\xi}\right)-
\varpi\left(-\frac{x}{\xi},-\frac{1}{\xi}\right)$, can be
restored, see Fig.\ \ref{FigParInt}. The uniqueness of this
continuation procedure was shown in connection with the support
extension of evolution kernels
\cite{GeyDitHorMueRob88MueRobGeyDitHor94}.

GPDs simultaneously possess a longitudinal and transversal momenta
dependence and so they encode the three dimensional distribution
of  partons in the considered hadron or nucleus \cite{RalPir01}.
Indeed, a partonic density interpretation holds in the infinite
momentum frame as long as $\xi=0$ \cite{Bur00Bur02}, see also
Refs.\ \cite{Die02,BelMue02}. In this kinematics the central
region, i.e., the parton number violating contributions, drops
out. Thus, it could be shown that in the impact parameter space
within the infinite momentum frame
\begin{eqnarray}
h_i(x,\mbox{\boldmath{$b$}}_\perp)=
\int\!\frac{d^2\mbox{\boldmath{$\Delta$}}_\perp}{(2\pi)^2}\, e^{i
\mbox{\boldmath{$b$}}_\perp\cdot\mbox{\boldmath{$\Delta$}}_\perp}\,
H_i(x,\xi=0,-\mbox{\boldmath{$\Delta$}}^2_\perp)
\end{eqnarray}
has the probabilistic interpretation to find a parton species $i$
in dependence on the momentum fraction $x$ and the relative
distance $\mbox{\boldmath{$b$}}_\perp$ from the proton center. It
is noted that an interpretation of the three dimensional Fourier
transform of GPDs in the rest frame has been suggested within the
concept of phase space (Wigner) distributions \cite{Bel03,Ji03}.

\section{Parametrization, model and lattice calculations of GPDs}
\label{Sec-ModLatCal}

GPDs are mostly unknown functions with a complex variable
dependence, which must satisfy the basic properties.  Because of
the lack of further knowledge, GPD ans\"atze, needed for the
estimation of cross sections and asymmetries, are constructed so
far by simplicity and intuition \cite{Rad98a}. Such an ansatz,
given at the input scale $\cQ_0^2$, is based on the assumption of
$(x,\xi)$ and $\Delta^2$ factorization and is now widely used in
phenomenology:
\begin{eqnarray}
\label{AnsGPD} H_i(x,\xi,\Delta^2,\cQ_0^2) = F_i(\Delta^2)
h_i(x,\xi)\,,\quad h_i(x,\xi) = \int_{-1}^1\, dy
\int_{-1+|y|}^{1-|y|} \delta(x-y -\xi\, z ) D_i(y,z)\,.
\end{eqnarray}
Here  $ F_i(\Delta^2)$ are partonic form factors, partially fixed
by sum rules. The reduced GPD $h_i(x,\xi)$ is given in terms of a
factorized ansatz for the so-called double distribution $D_i(y,z)=
q_i(y\cQ_0^2) \Pi(|z|/(1-|y|))/(1-|y|) $ with the parton density
$q_i(y,\cQ_0^2)$ and an unknown profile function $\Pi$
\cite{Rad98a,GoePolVan01}. This ansatz might be refined by a
regge-like ansatz for the parton densities at small $x$:
$q_i(x)\sim x^{-\alpha_i} \to x^{-\alpha_i - \alpha_i^\prime
\Delta^2} $ \cite{GoePolVan01}. To repair an artifact in the
relation among GPD and DD \cite{BelMueKirSch00Ter01}  a so-called
$D$-term is added in (\ref{AnsGPD}), intuitively understood as
meson exchange contribution in the $t$-channel \cite{PolWei99}.

In the region $\xi < x$ for small $x$ it also has been proposed to
equate the reduced GPDs with the parton densities at the input
scale and so skewness is purely generated by evolution
\cite{FraFreGuzStr97}. We note that in this region, where  gluon
and sea quark contributions are dominating, the parameterization
of the hard-scattering amplitude convoluted with a GPD drastically
simplifies. Here, the skewness effect at the input scale
effectively enters the normalization \cite{BelMueKir01}. A further
proposal for the GPD parametrization is based on their
representation as an infinite sum of $t$-channel exchange
contributions \cite{PolShu02}.

The GPD parameterizations, proposed so far, do not simultaneously
satisfy the needs:
\begin{itemize}
\item Basic properties must be automatically fulfilled. \item
Flexible parametrization of the degrees of freedom that are left.
\item Simple numerical treatment of evolution and convolutions.
\end{itemize}
These requests can be mostly satisfied within a generalization of
the Mellin representation for parton densities by adopting the
concept of complex angular momentum to the conformal spin
expansion of GPDs \cite{MueSch05}.

The poor knowledge about GPDs might be improved by
non-perturbative model calculations, done at a low $\mu^2$ scale.
The resulting GPDs can then be evolved to the scale that is
relevant for their phenomenological use. However, the scheme
dependence of GPDs leads to uncertainties in this matching
procedure, which should be considered as  part of the model.
Because of limited space, all the efforts cannot be  reviewed here
in detail, however, at least it should be mentioned a few of them,
namely, calculations within the MIT bag model \cite{JiMelSon97},
chiral quark soliton model \cite{PetPobPolBoeGoeWei97},
constituent quark model \cite{ScoVen02BofPasTra02}, and
Bethe-Salpeter and light-cone wave function approaches
\cite{ChoJiKis01TibMil02NogTheVen02}.

More recently, the first few Mellin moments (\ref{Def-MellMom})
for proton GPDs have been measured on the lattice
\cite{Hagetal03Gocetal03}. Especially, the quark orbital angular
momentum fraction of the proton spin, the second moment of a
certain GPD combination, could be extracted. Also the correlation
of the $\Delta^2$ dependence with the order $n$ of moments leads
to a valuable insight in the nucleon GPDs. The slope of the
$\Delta^2$ dependence decreases with increasing $n$ and so  the
transversal size of the proton shrinks with increasing $x$, which
has been argued in Ref.\ \cite{Bur00Bur02}. From this one
concludes that the factorized ansatz (\ref{AnsGPD}) is
oversimplified.  It might be also possible to rich from lattice
results a better understanding of the skewness dependence. So far
the common GPD ans\"atze leads always to an enhancement of the
(reduced) GPDs, compared to parton densities, at the crossing
point $x=\xi$. For the phenomenology it is highly desired to have
a more rigorous understanding of this issue.

\section{Hard exclusive reactions to leading order}
\label{Sec-ExcReaLO}

\begin{figure}[t]
\unitlength1mm
\begin{minipage}[hb]{180mm}
\begin{minipage}[hb]{60mm}
\begin{picture}(80,20)(0,0)
\put(10,-8){\insertfig{6}{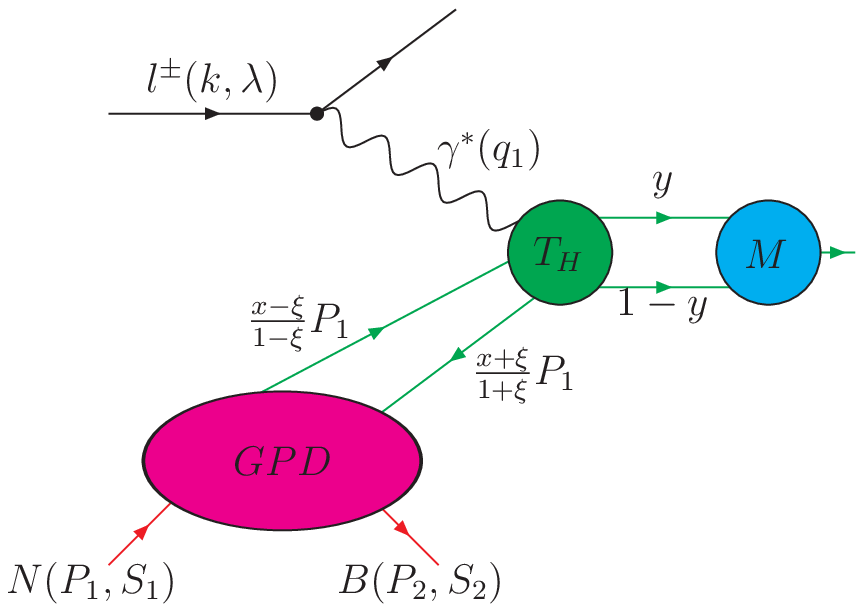}}
\end{picture}
\end{minipage}
\hfill
\begin{minipage}[hb]{80mm}
\vspace{-5mm} \noindent
Hadronic variables:\\
$\cQ^2= -q_1^2$,\\ $
\Bx =  \frac{\cQ^2}{2P_1\cdot q} \approx \frac{2 \xi}{1+\xi}$\\
{\ }\\
Scaling limit:\\
$\cQ\to \infty,\ \Bx$ -- fix \\
$\Delta^2 = (P_2-P_1)^2 $ -- fix
\end{minipage}
\end{minipage}
\caption{ \label{Fig-HarMesLepPro} Factorization of the meson
leptoproduction amplitude (left) and kinematics.}
\end{figure}
GPDs appear in the perturbative description of the hard photon or
meson leptoproduction: $l^\pm N \to l^\pm B X$, where $B$ is the
final baryon state and $X$ stands for the photon or the observed
meson. The virtuality $- \cQ^2$ of the intermediate photon must be
sufficiently large. This implies that  the hadronic scattering
amplitude factorizes into a hard-scattering one convoluted with
GPDs and eventually a meson distribution amplitude (DA), see Fig.\
\ref{Fig-HarMesLepPro}
\cite{ColFraStr96,ColFre98Fre99b,Rad97,JiOsb98}. Usually, one
defines this factorization in such a way that the collinear
divergencies are removed within a (modified) minimal subtraction
 scheme.  Obviously, this procedure is
ambiguous and induces, e.g., the factorization scale  dependence.
If the same factorization scheme is applied for all processes,
GPDs and DAs are universal in the sense that they are process
independent, however,  they depend on the scheme and of the order
at which the perturbation theory is truncated. Thus,  GPDs serve
at the first place as a tool that {\it connect} physical
observables measured in different processes. Concerning their
probabilistic interpretation, one must conclude that the number of
measured partons depends on the scheme conventions (even in
untruncated perturbation theory). This problem, appearing also for
parton densities, might be resolved by choosing a reference
process, e.g., the leptoproduction of photons, in which all
radiative corrections (order by order) are absorbed in the GPD
definitions.

Let us first consider the leptoproduction of a meson. So far the
quantum numbers allow such a process and the virtual photon is
longitudinally polarized, the process is for sufficient large
$\cQ^2$ perturbatively described as shown in Fig.\
\ref{Fig-HarMesLepPro}. The initial and final hadron state might
have different quantum numbers and so it is applicable for several
processes, namely, the production of neutral and charged vector
\cite{Hoo96ManPilWei97aManPilWei97} and pseudo scalar
\cite{VanGuiGui98ManPilRad98EidFraStr98FraPolStrVan99} mesons and
or even for exotic states like the pentaquark \cite{DiePirSzy04}
or hybrid mesons \cite{AniPirSzyTerWal04}. Besides the cross
section also the transversal target spin asymmetry is described
within the perturbative framework, which might be justified at a
lower scale $\cQ^2$ as for the cross section itself
\cite{FraPobPolStr99,BelMue01a}. To LO the convolution of GPDs and
meson DA are separated
\begin{eqnarray}
\label{Pre-Pro}
A_L(\xi,\Delta^2,\cQ^2,S_1,S_2)\!\!\! &\propto &\!\!\!\!\! \frac{\alpha_s}\cQ
\sum_{f,\bar{f}=u,\dots,g} \!\int_{-1}^1\! dx\! \int_0^1\! dy \;
q^{f\bar{f}}(x,\xi,\Delta^2,\cQ^2,S_1,S_2)
\nonumber\\
&&\times\Bigg[\frac{Q_f}{(1-y)(x-\xi+i\epsilon)}
+ \frac{Q_{\bar f}}{y (x + \xi-i\epsilon)} \Bigg]
 \phi^{f\bar{f}}_M(y,\cQ^2)\,
\nonumber
\end{eqnarray}
where $q^{f\bar{f}}$ and $\phi^{f\bar{f}}_M$ refers to the
(transition) GPD and meson DA, respectively. $ Q_f, Q_{\bar{f}}$
are the partonic charge factors, corresponding to the flavor
content of the observed meson. Hence, the produced meson  serves
as a flavor filter and so a flavor decomposition for GPDs can be
reached, e.g., by measuring the processes  $e^- p \to e^- p \pi^0$
and $e^- p \to e^- p \eta$. The amplitude $A_L$ scales with
$1/\cQ$, which leads to a scaling law $d\sigma/d\Delta^2 \propto
1/\cQ^6$ for the cross section. However, this canonical scaling is
logarithmically modified by evolution effects, mainly arising from
GPDs.

Several of these processes have been  measured or a planned in
fixed target (HERMES, JLAB, COMPASS) and collider (H1, ZEUS)
experiments, for a review see the contributions of F.~Sabatie and
L.~Favart in this proceedings. I only like to give a short
attention to the electroproduction of neutral (longitudinally
polarized) vector mesons, measured in the small $x_{\rm Bj}\sim
2\xi$ region up to rather large $\cQ^2$ by the H1 and ZEUS
collaborations. Here the amplitude (\ref{Pre-Pro}) drastically
simplifies, since now it is dominated by gluon exchange, and thus
such experiments are an ideal testing ground of the perturbative
framework and the study of the gluonic GPD. Let me remind that
under the assumption of SU(3) symmetric meson DAs the ratio of
cross sections should be:
$\sigma_{\rho^0}:\sigma_{\omega}:\sigma_{\phi} = 9:1:2$, which is
in fair agreement with experimental data (plotted with respect to
the scale $\cQ^2 + M^2_V$).

\begin{figure}[t]
\begin{center}
\vspace{0cm}
\hspace{0cm} \mbox{
\begin{picture}(600,25)(0,0)
\put(-10,-25){\insertfig{7}{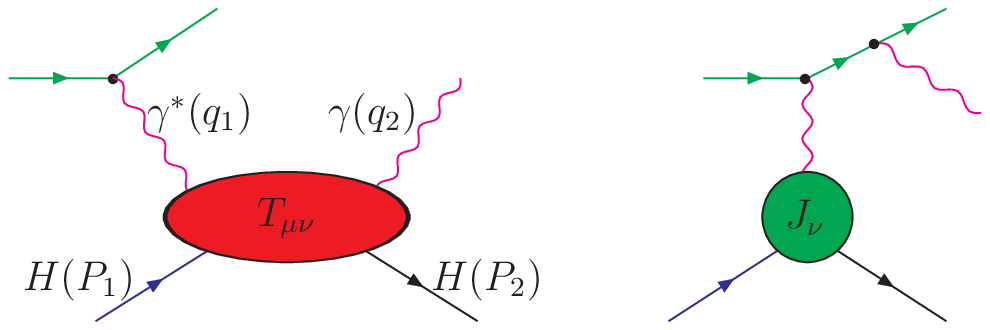}}
\put(250,-25){\insertfig{7}{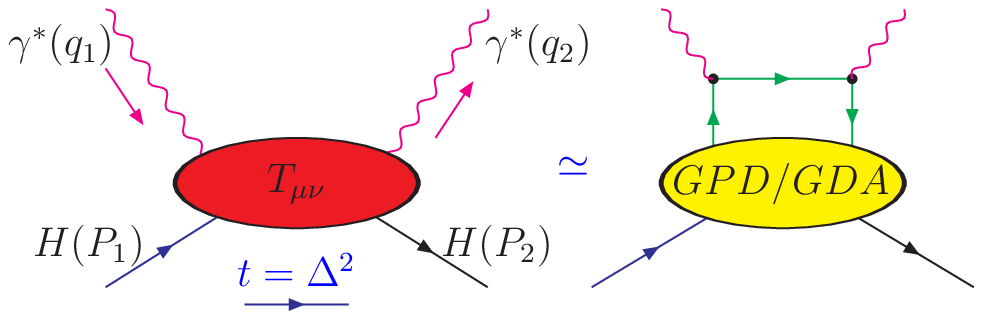}}
\end{picture}
}
\end{center}
\caption{ \label{CSpro} Deeply virtual Compton scattering and
Bethe-Heitler bremsstrahlungs process (left) and factorization of
the Compton amplitude to leading order (right).}
\end{figure}
Another class of hard exclusive processes, in which GPDs are
accessible, are those in which the struck parton is purely probed
within the electromagnetic interaction via the absorbtion and
emission of photons \cite{GeyDitHorMueRob88MueRobGeyDitHor94}:
photo leptoproduction \cite{DieGouPirRal97,BelMueKir01}, see Fig.\
\ref{CSpro}, lepton pair photoproduction \cite{BerDiePir01} and
lepton pair leptoproduction \cite{GuiVan02BelMue02aBelMue03}. In
the first two processes the photon virtuality is space- and
time-like respectively, while in the third one both photons are
virtual. The latter one is experimentally most challenging,
however, it is the only one in which a deconvolution of the GPDs
is possible. In all these processes there are two interfering
subprocesses, the hard virtual Compton scattering and the
Bethe-Heitler (BH) bremsstrahlungs process, given in terms of the
electromagnetic form factors. For the case that at least one
photon has a large virtuality, the Compton scattering process
factorizes, shown in Fig.\ \ref{CSpro} to LO.

Hence, the cross section has a rich azimuthal angular dependence
and the interference term is linear in the GPDs. This offers the
possibility to unreveal the GPDs by measuring asymmetries that are
dominated by the interference term: charge, single beam and target
spin asymmetries, while double spin asymmetries require the
subtraction of the BH contribution. In collider experiments it is
even possible to extract the deeply virtual Compton scattering
(DVCS) cross section, since the BH amplitude is sufficiently
small. A detailed compendium of asymmetries and their relations to
GPDs is given in Ref.\ \cite{BelMueKir01}. The first measurements
of single beam spin and charge asymmetries and cross sections are
compatible with the oversimplified GPD ansatz (\ref{AnsGPD}).

It was also argued that the amplitudes of the photon and meson
photoproduction at large $s$ and $-t$ factorize to LO in terms of
the inverse  GPD moments $\int_{-1}^{1}\! dx\, q(x,\xi,t)/x$
\cite{Rad98c,DieFelJakKro98,HuaKro00}. For these processes also a
perturbative treatment is used in which all valence partons are
resolved via hard gluon exchanges \cite{BroLep80}. In this
conjecture here only the struck parton is resolved while all
spectators are contained in the GPD. This is somehow analogous to
the mechanism proposed by Feynman for the description of elastic
form factors. Unfortunately, no factorization theorem could be
established, since power corrections are uncontrolled. This
problem has been attempted to resolve by the assumption that the
virtuality of the partons in the initial and final state is always
small \cite{DieFelJakKro98}.

Note that the perturbative framework is also applicable for the
photon \cite{KirMue02KirMue03CanPir02GuzStr03} and meson
\cite{CanPir03} leptoproduction off nuclei. This opens a new
window for the study of nucleus binding effects, especially, for
the deuteron \cite{BerCanDiePir01,Pol02}.

\section{Beyond leading order predictions}
\label{Sec-BeyLO}

To gain insight in the validity of the perturbative framework and
to improve it, it is necessary to calculate  higher order
perturbative and power suppressed corrections or at least to
estimate them. A large amount of work has been done in this
direction during the last few years.

The factorization theorems, derived to leading power accuracy,
state that the perturbative corrections are systematically
calculable in this approximation. The next-to-leading order (NLO)
corrections for the hard-scattering amplitude  of both Compton
scattering processes \cite{BelMue97aManPilSteVanWei97,JiOsb98} and
for leptoproduction of mesons have been completed
\cite{BelMue01a,IvaSzyKra04}. While for the former process the
perturbative corrections, which depends on the GPD ansatz, turn
out to be moderate, the size of NLO corrections for the latter is
rather large. However, these corrections partly cancel in the
transversal proton spin asymmetry \cite{BelMue01a}.

Employing conformal consistency predictions \cite{Mue94BelMue98c},
the evolution kernels have been evaluated to NLO
\cite{BelFreMue99} and are implemented in numerical codes
\cite{FreMcD01a}. Note that conformal symmetry in the minimal
subtraction scheme is broken in a subtle manner, which can be
removed by a finite renormalization, providing the conformal
subtraction (CS) scheme \cite{Mue97a}. In such a scheme the
next-to-next-to-leading order (NNLO) corrections to both the DVCS
hard-scattering amplitude and evolution kernels can be borrowed
from the known results in deep inelastic scattering, for a first
discussion see Ref.\ \cite{MelMuePas02}.

Whether power suppressed contributions can be calculated within
perturbative QCD depends on the process in question. For hard
exclusive leptoproduction the exchange of a transversal polarized
photon yields  $1/{\cQ}$ suppressed contributions. However, they
are affected by non-integrable singularities, which appear to LO
in $\alpha_s$ and are induced by large size quark-antiquark
configurations of the produced meson \cite{ColFraStr96}. In the
case of photon leptoproduction $1/\cQ$ suppressed contributions
are calculable to LO \cite{BelMue00a} and NLO \cite{KivMan03}
accuracy. In fact, they complete the azimuthal angular dependence
of the cross section \cite{BelMueKir01}. Interesting to remark,
that within an appropriate definition of observables  they do not
interfere with the twist-two prediction, appearing at leading
power.

Twist-two predictions will be affected by $1/\cQ^2 $ power
suppressed contributions appearing at twist-four level. Such
contributions have been perturbatively calculated to LO by
neglecting multi-particle operators
\cite{BelMue01KivMan01EilGeyRob04}. Unfortunately, this
approximation suffers from an ambiguity in the choice of the
multi-particle operator basis, which shows up in the violation of
current conservation. This problem is not resolved so far.  Upper
bounds of the power corrections to the hard electroproduction have
been estimated within the renormalon approach and found to be
large \cite{Bel03}.

The fact that perturbative and non-perturbative corrections for
leptoproduction of mesons are much larger as for photons indicates
that the onset of the scaling region for the former process will
be at a higher scale as for the latter one. For photon
leptoproduction one would expect that, as in the case of deep
inelastic scattering, a scale of few $\GeV^2$ is sufficient to
apply perturbative QCD.

\section{Conclusions}

In the last decade the concept of generalized parton distributions
and distribution amplitudes has been enormously developed: a
partonic interpretation has been given in depth,  factorization
theorems has been derived, predictions for numerous processes has
been worked out, radiative corrections are calculated to NLO
accuracy, including the evolution equations, first measurements on
the lattice has been performed, and experimental accessibility has
been demonstrated in pioneering experiments.

GPDs and also GDAs  can be explored within measurements of hard
exclusive, however, inelastic processes in fixed target and
collider experiments, where a deconvolution with respect to the
momentum fraction is practically impossible. Only the
leptoproduction of a lepton pairs allows to scan the central
region of GPDs. These functions are hybrids that incorporate
parton densities, form factors, and distribution amplitudes. Thus,
they form a link between exclusive and inclusive processes and
encode information about non-perturbative QCD, which cannot be
obtained from inclusive or elastic exclusive processes. GPDs/GDAs
are a new tool that allows to study the structure of hadrons and
nuclei from a new perspective.  So for instance, their knowledge
would provide the quark orbital angular momentum contributing to
the hadron spin and the three dimensional distribution of partons
inside hadrons and nuclei. Moreover, they can serve for the study
of flavor and chiral symmetry breaking or nucleus binding effects,
to name a few. As it has been stressed, GPDs are scheme dependent
quantities and, thus, their probabilistic interpretation should be
done within an appropriate scheme convention.

The solution of several open problems will require a large effort.
The most challenging issue for the theory is whether factorization
is applicable for the photon virtuality reached in present
experiments. Here higher order calculations beyond the NLO
accuracy might give some insight. Also the growing amount of
experimental data will help to answer this question. A further
issue concerns the appropriate and realistic parametrization of
GPDs. From the first photon leptoproduction measurements it
becomes also clear that high precision data are needed to
distinguish between GPD models. Certainly, in the last decade a
huge, however, first step has been taken in reaching a deeper
understanding of the hadron and nuclei structure within the
concept of GPDs/GDAs.


\end{document}